\documentclass[preprint,showpacs,
preprintnumbers,amsmath,amssymb,showkeys]{revtex4}
\usepackage{graphicx}
\usepackage{dcolumn}
\usepackage{bm}
\usepackage[usenames]{color}
\usepackage[latin1]{inputenc}
\usepackage{fancyhdr}
\usepackage{verbatim} 
\usepackage{cancel}
\usepackage{bm}
\usepackage{amsmath,amssymb,amsthm,amscd,mathrsfs}
\usepackage{rawfonts}
\usepackage{pictexwd}
\usepackage{nccmath}
\usepackage{multirow,array}
\usepackage{float}
\usepackage{braket}
\usepackage{color}
\usepackage{endnotes}
\usepackage{tabulary} 
\usepackage{booktabs}
\usepackage{graphicx}
\usepackage{bm}
\usepackage{subfigure} 
\DeclareGraphicsExtensions{.pdf,.png,.jpg}

\begin{document}
\title{On the analytical solutions of the quasi-exactly solvable Razavy type potential
\bm{ $V(x)=V_0\left( {\rm sinh}^4(x)- k\, {\rm sinh}^2(x) \right)$} }
\author{Marco A. Reyes}
\email{marco@fisica.ugto.mx}
\affiliation{Divisi\'on de Ciencias e Ingenierías, Universidad de Guanajuato, Guanajuato, M\'exico}
\author{Edgar Condori-Pozo}
\affiliation{Divisi\'on de Ciencias e Ingenierías, Universidad de Guanajuato, Guanajuato, M\'exico}
\author{Carlos Villase\~nor-Mora}
\affiliation{Divisi\'on de Ciencias e Ingenierías, Universidad de Guanajuato, Guanajuato, M\'exico}
\author{M. Ranferi Guti\'errez}
\affiliation{Instituto de Investigaciones de Ingenier\'ia, Matem\'aticas y Ciencias F\'{\i}sicas, Universidad Mariano Galves, Guatemala, Guatemala}
\date{May 2018}
\begin{abstract}
In order to show how stringent the restrictions posed on analytical solutions of quasi-exactly solvable potentials are, we construct analytical solutions for the Razavy type potential 
$V(x)=V_0\left( {\rm sinh}^4(x)- k\, {\rm sinh}^2(x) \right)$ based on the polynomial solutions of the related Confluent Heun Equation, where the free parameter $k$ allows to tune energy eigenvalues, a desirable feature in different theories.  However, we show that with the described method, the energy eigenvalues found diverge when $k\to -1$, a feature caused solely by the procedure.

\end{abstract}



\maketitle


\section{Introduction}

In quantum mechanics, an exactly solvable problem is that where the whole spectrum is found analytically.  While the vast majority of problems have to be solved numerically, 
a new possibility arised with the class of so called QES potentials, where a subset of the whole spectrum may be found analitycally.\cite{Turbiner,Shifman,Ushveridze1}

Manning,\cite{Qiong} Razavy\cite{Razavy}, and Ushveridze\cite{Ushveridze2} type potentials belong to this class
(see also \cite{Chennn}). 
They also belong to the class of double well potentials, which have received much attention due to their applications in theoretical and experimental problems. Furhermore, hyperbolic type  potentials are found in many physical applications, like the Rosen-Morse potential,\cite{Oyewumi} Dirac type hyperbolic potentials,\cite{Wei} bidimensional quantum dot,\cite{Xie} Scarf type entangled states,\cite{Downing} etc.
QES potentials classification have beens given by Turbiner,\cite{Turbiner} and Ushveridze.\cite{Ushveridze2}

In this article we construct analytical solutions for the Razavy type potential 
$V(x)=V_0\left( {\rm sinh}^4(x)- k\, {\rm sinh}^2(x) \right)$, 
which for $k>0$ may be seen as the infinite well counterpart of Manning's, 
in terms of the polynomial solutions of the confluent Heun equations (CHE) associated to the problem.  
As with other QES potentials, we will show that the solutions, or in this case the parameter $V_0$,  depend on the order of the polynomial used to solve it.  We also show that 
the energy eigenvalues diverge as $(1+k)^{-1}$ when $k\to -1$, but this is only a characteristic due to the analytical solution method.

The article is organized as follows. In Section \ref{genral} we relate our potential function to those of Razavy and Ushveridze, and in Sections \ref{seckeq0} and \ref{seckeq0b} we describe the general equation transformations that lead to CHEs for both even and odd solutions in the case of $k=0$.  In Section \ref{seckgt0} we find solutions for various cases of the double well potential, with $k>0$, and in Section \ref{secnew} we describe the singular aspects of the analytical solutions.
Section \ref{final} contains our final remarks. 


\section{Relation to Razavy and Ushveridze potentials}\label{genral}

We consider the following Schr\"odinger's equation
\begin{equation} \label{ecuaciondeschrodinger}
\frac{-\hbar^2}{2m} \frac{d^2 \psi(x)}{d x^2} +  
V_0 \left( \sinh^{4}(\lambda x) -k\, {\rm sinh}^2(\lambda x) \right) \, 
\psi(x)=E \, \psi(x)
\end{equation}
where $\lambda$ is a constant used to fix dimensions. For simplicity, we set $m=\hbar=\lambda=1$.
To solve this equation we will use as reference the works of Downing {\it et al.},\cite{Downing} and Wen {\it et al.}\cite{Wen}

The potential function in eq.(\ref{ecuaciondeschrodinger}) may be seen as the hyperbolic Razavy potential 
$V(x)=\frac{1}{2}\left( \zeta \,{\rm cosh}(2x)-M \right)^2$, with $V_0=2\zeta^2$, for which Razavy showed that if $M$ is a positive integer, one can compute exactly the first $M$ energy levels.\cite{Razavy}  It may also be viewed as a particular case of the Ushveridze potential 
$V(x)=2\xi^2 \,{\rm sinh}^4(x)+2\xi\left[ \xi-2(\gamma+\delta)-2\ell \right] {\rm sinh}^2(x)
+2(\delta-\frac{1}{4})(\delta-\frac{3}{4})\, {\rm csch}^2(x)
-2(\gamma-\frac{1}{4})(\gamma-\frac{3}{4})\, {\rm sech}^2(x)$,
when $\gamma=\frac{1}{4}$ and $\delta=\frac{3}{4}$, or viceversa.\cite{Ushveridze2} In this case also $V_0=2\xi^2$.  Ushveridze showed that this potential is QES if $\ell=0,1,2,\cdots$ (with $\delta\ge \frac{1}{4}$), and  
El-Jaick {\it et al.} showed that it is also QES if $\ell=$half-integer and
$\gamma,\delta=\frac{1}{4},\frac{3}{4}$,\cite{Jaick} which is the potential we are considering.
Note that in both cases the constants $M$ and $\ell$ are fixed to be integers or half-integers in order to relate the solutions to the polynomial solutions of the CHEs involved.

In the case of the Razavy potential, the solutions obtained by Finkel {\it et al.}, are
\begin{equation}
\psi_{\sigma \eta} \left( x,E_{R} \right) \propto \left( \sinh x \right)^{ \frac{1}{2} \left( 1 - \sigma - \eta \right)} \left( \cosh x \right)^{\frac{1}{2} \left( 1 - \sigma + \eta \right)} e^{- \frac{\zeta}{2} \cosh(2x)} \sum_{j=0}^{n} \frac{\hat{P}_{j}^{\sigma \eta} \left(E_{R} \right)}{ \left( 2j + \frac{\eta - \sigma + 1}{2} \right) !} \cosh^{2j}(x)
\end{equation}
with the parameters $(\sigma,\eta)=(\pm 1,0)$ or $(0,\pm 1)$, the energy eigenvalues being the roots of the polynomials $P_{j+1}^{\sigma \eta}(E_R)$, satisfying the three term recursive relations
\begin{equation}
\hat{P}_{j+1}^{\sigma \eta} = \left( E_{R} - b_{j} \right) \hat{P}_{j}^{\sigma \eta} \left( E_{R} \right) - a_{j} \hat{P}_{j-1}^{\sigma \eta} \left( E_{R} \right), \qquad j \geq 0
\end{equation}
with $E_R=2E$, and
\begin{equation}\label{ajbj}
\begin{matrix}
a_j=16\zeta j(2j-\sigma+\eta)(j-n-1)
\\
b_{j} = -4j \left( j + 1 - \sigma + 2 \zeta \right) + \left( 2n + 1 \right) \left( 2 \left( n - \sigma \right) + 3 \right) + \zeta \left( \zeta - 2 \eta + 4n \right)
\end{matrix}
\end{equation}

In this article we shall simply construct solutions of eq.(\ref{ecuaciondeschrodinger}) based on polynomial solutions of CHEs derived from the change of variable considered.


\section{Symmetric solutions for $\boldsymbol{V(x) = V_0 ~ \mathrm{sinh}^{4}(x)}$}\label{seckeq0}

To begin with, let us look for the solutions for the case $k=0$, and return to the case with $k>0$ 
in Section \ref{seckgt0}.
Since we are considering a continuous parameter $k$, we can think of the case with $k=0$ as the double to single well deformation of the case with $k>0$.  

In order to find the even solutions to eq.(\ref{ecuaciondeschrodinger}), let us change the independent variable to $\beta(x) = \cosh^2(x)$, to get the equation
\begin{equation} \label{ecuacion4}
\beta \left( \beta -1 \right) \frac{d^{2} \psi}{d \beta^{2}} + \left( \beta - \frac{1}{2} \right) \frac{d \psi}{d \beta} + \frac{1}{4} \left[ 2 E - 2 V_0 \beta^{2} + 4 V_0 \beta - 2 V_0 \right] = 0
\end{equation}
and then, to ensure that $\psi (x)$ vanishes as $x\to\pm\infty$, propose that 
$\psi \left( x \right) = e^{-\frac{\alpha}{2} \beta} f(\beta)$.  Note that previous works may not include square integrable polynomal solutions to the Razavy potential.\cite{no2int2,no2int3,no2int}
By requiring $\alpha^{2} = 2 V_{0}$, we obtain the following CHE
\begin{equation} \label{ecuacion5.1}
\beta \left( \beta - 1\right) \frac{d^{2} f}{d \beta^{2}} + \left[ - \alpha \beta \left( \beta -1 \right) + \left( \beta - \frac{1}{2} \right) \right] \frac{d f}{d \beta} + \left[ \frac{\alpha^{2} \beta}{4} - \frac{\alpha \beta}{2} + \frac{\alpha}{4} + \frac{E}{2} - \frac{\alpha^{2}}{4} \right] f = 0 
\end{equation}
which has finite series solutions.\cite{Yao}

Now, instead of simply looking for Heun function solutions to eq.(\ref{ecuacion5.1}), 
we shall 
directly look for rank $N$ polynomial solutions: $f(\beta)$=$f_0$ for $N=0$, 
or $f(\beta)$=$f_0 \prod_{i=1}^{N} \left( \beta - \beta_i \right)$ for $N>0$, the $\beta_i$ being the roots of the resulting polynomial in eq.(\ref{ecuacion5.1}).  Note that the solution with $N$=$0$ sometimes is not considered among these solutions.\cite{Downing}

Given this, it turns out that the highest powers of $\beta$ in eq.(\ref{ecuacion5.1})
fix $\alpha$ to be $\alpha = 4 N + 2$, and the energy eigenvalues and the roots satisfy
\begin{equation}
E = \frac{1}{2} \left[ \alpha^{2} + \alpha \left( 4 \sum_{i=1}^{N} \beta_{i} - 1 - 4 N \right) - 4 N^{2} \right] \label{ecuacion2.101c1}
\end{equation}
\begin{equation} \label{ecuacion2.102c1}
\sum_{i \neq j}^{N} \frac{2}{\beta_{i} - \beta_{j}} + \frac{-\alpha \beta_{i}^{2} + \left( \alpha + 1 \right) \beta_{i} - \frac{1}{2}}{ \beta_{i}^{2} - \beta_{i}} = 0, \qquad i = 1,2,\ldots, n
\end{equation}
The parameter $V_0$ is found to depend on the order of the polynomial, with  
$V_{0} = 2(2N+1)^2$ for all even solutions, and solutions with different $N$ values can not be scaled one into the other due to the $\rm{sinh}^4(x)$ dependence of the potential function.  Also, the maximum order $n$ of eigenfunctions is given by $n=2N$.
We shall give here the results for $N=0$ and 2.\cite{edgar}
We use subindexes $\left\{N,n\right\}$ to label eigenvalues/eigenfunctions, the polynomial and eigenfunctions orders.


As an example, let us consider the case with $N=0$. Setting $f(\beta)=1$, 
we find that $V_0 = 2$, and $E_{0,0} = 1$ for the (unnormalized) ground state eigenfunction
$\psi_{0,0} \left( x \right) = e^{- \cosh^{2} \left( x \right)}$. 
In the case of higher order polynomial solutions, let us consider the case of $N = 2$, $f(\beta)=f_0(\beta-\beta_1)(\beta- \beta_2 )$.
Equating to zero the coefficients of different powers of the polynomial in $\beta$, we get that
\begin{equation}\label{eq7}
\begin{matrix}
&\frac{\alpha^{2}}{4} - \frac{5 \alpha}{2} = 0 \\
&3+ \left( \beta_1 + \beta_2 \right) \left( - \frac{\alpha^{2}}{4} + \frac{3\alpha}{2} \right) + \left( - \frac{\alpha^{2}}{4} + \frac{9 \alpha}{2} + \frac{E}{2}\right) = 0 \\
&-3 - \left( \beta_1 + \beta_2 \right) \left( - \frac{\alpha^{2}}{4} + \frac{5 \alpha}{4} + \frac{E}{2} + 1 \right) + \beta_1 \beta_2 \left( \frac{\alpha^{2}}{4} - \frac{\alpha}{2} \right) = 0 \\
&\frac{1}{2} \left( \beta_1 + \beta_2 \right) + \beta_1 \beta_2 \left( -\frac{\alpha^{2}}{4} + \frac{\alpha}{4} + \frac{E}{2} \right) = 0
\end{matrix}
\end{equation}
Solving these, we find that $V_{0} = 50$, and there are 3 possible eigenvalues and eigenfunctions
\begin{equation}
\begin{matrix}
E_{2,0} = \ 2.6301, & 
\psi_{2,0} = e^{-5 \cosh^{2} \left( x \right)} \left( \cosh^{2} \left( x \right) - 0.0260 \right) \left( \cosh^{2} \left( x \right) - 0.2555 \right) \\
E_{2,2} = 19.0121, &  
\psi_{2,2} = e^{-5 \cosh^{2} \left( x \right)} \left( \cosh^{2} \left( x \right) - 0.0401 \right) \left( \cosh^{2} \left( x \right) - 1.0659 \right) \\
E_{2,4} = 43.2490, &  
\psi_{2,4} = e^{-5 \cosh^{2} \left( x \right)} \left( \cosh^{2} \left( x \right) - 1.0288 \right) \left( \cosh^{2} \left( x \right) - 1.2836 \right)
\end{matrix}
\end{equation}

\begin{figure}[H]
\centering
{\includegraphics[width=0.7\textwidth]{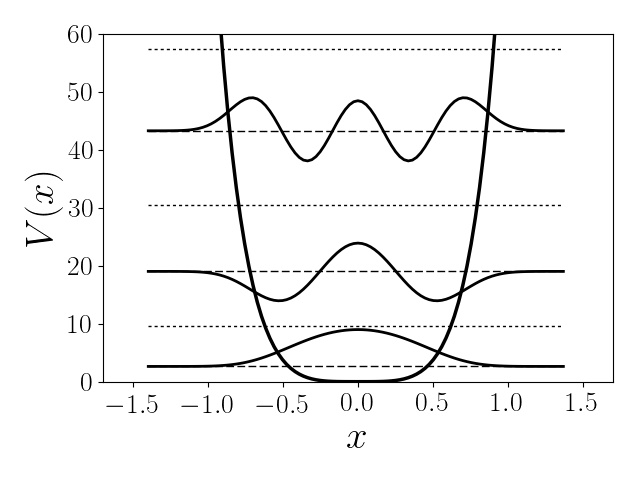}} 
\caption{The three even eigenfunctions (narrow solid lines) found analytically for $N = 2$, with $V_0=50$, together with the corresponding eigenvalues (dashed lines).  The unsolved odd eigenvalues, found using Numerov's method, are shown in dotted lines.}
\label{figuno}
\end{figure}

\noindent 
The unnormalized eigenfunctions are plotted in Fig.(\ref{figuno}).  Since the polynomial in $\beta(x)$ is $O(2)$, there are three posible solutions to eq.(\ref{eq7}), each giving two roots $\beta_i$ of eq.(\ref{ecuacion5.1}), which define the three eigenfunctions with 0, 1 or 2 nodes, depending if these roots are $\beta_i>1$.


\section{Antisymmetric solutions}\label{seckeq0b}

In order to find antisymmetric solutions to eq.(\ref{ecuacion5.1}), we only need to do the substitution
$f(\beta) = \mathrm{sinh}(x) \, g(\beta)$, to obtain
\begin{align} \label{ecuacion32} 
\nonumber \beta \left[ \beta - 1 \right] \frac{d^{2}g}{dx^{2}}  &+  \left[ - \alpha \beta^{2} +  \left( \alpha + 2 \right) \beta - \frac{1}{2} \right] \frac{d g}{dx} \\
&+ \left[ \left( - \alpha + \frac{\alpha^{2}}{4} \right) \beta + \left( - \frac{\alpha^{2}}{4} + \frac{\alpha}{4} + \frac{E}{2} + \frac{1}{4} \right) \right] g = 0 
\end{align}
which is another CHE, and can be solved in power series: 
$g(\beta) = g_{0}$ if $N = 0$, or  $g(\beta)=g_{0}\, \prod_{i=1}^{N} \left( \beta - \beta_i \right)$ for $N > 0$, 
the $\beta_i$ being the roots of the resulting polynomial in eq.(\ref{ecuacion32}).  
Given this, it turns out that $\alpha = 4 (N + 1)$, and 
\begin{equation}
E = \frac{1}{2} \left[ \alpha^{2} + \alpha \left( 4 \sum_{i=1}^{N} \beta_{i} - 1 - 4 N \right)
 - 4 N^{2} -4N -1 \right] \label{ecuacion2.101c2}
\end{equation}
For the odd solutions, $V_{0} = 8(N+1)^2$, meaning that all even and odd solutions have different $V_0$.  The maximum order $n$ of eigenfunctions is $n=2N+1$.


For example, for $N = 3$, using $g = g_0 \left( \beta - \beta_1 \right) \left( \beta - \beta_2 \right)  \left( \beta - \beta_3 \right)$, we get that $\alpha=16$, and
\begin{equation}
\begin{matrix}
&\left( \beta_1 + \beta_2 + \beta_3 \right) \left( 3 \alpha - \frac{\alpha^{2}}{4} \right) + \left( - \frac{\alpha^{2}}{4} + \frac{13 \alpha}{4} + \frac{E}{2} - \frac{49}{4}\right) = 0 \\
&\left( \beta_1 + \beta_2 + \beta_3 \right) \left( \frac{\alpha^{2}}{4} - \frac{9 \alpha}{4} - \frac{E}{2} - \frac{25}{4} \right) + \left( \beta_1 \beta_2 + \beta_2 \beta_3 + \beta_3 \beta_1 \right) \left( \frac{\alpha^{2}}{4} - 2 \alpha \right) - \frac{15}{2} = 0 \\
&3 \left( \beta_1 + \beta_2 + \beta_3 \right) + \left( \beta_1 \beta_2 + \beta_2 \beta_3 + \beta_3 \beta_1 \right) \left( -\frac{\alpha^{2}}{4} + \frac{5 \alpha}{4} + \frac{9}{4} + \frac{E}{2} \right) + \beta_1 \beta_2 \beta_3 \left( - \frac{\alpha^{2}}{4} + \alpha \right) = 0 \\
&- \frac{1}{2} \left( \beta_1 \beta_2 + \beta_2 \beta_3 + \beta_3 \beta_1 \right) - \beta_1 \beta_2 \beta_3 \left( \frac{\alpha^{2}}{4} - \frac{\alpha}{4} - \frac{E}{2} - \frac{1}{4} \right) = 0
\end{matrix}
\end{equation}
with $V_{0} = 128$.  We find four eigenvalues, 
$E_{3,1} = 12.8152$, 
$E_{3,3} = 40.4568$,   
$E_{3,5} = 75.7246$, and   
$E_{3,7} = 117.003$,
for the eigenfunctions

\begin{figure}[H]
\centering
{\includegraphics[width=0.7\textwidth]{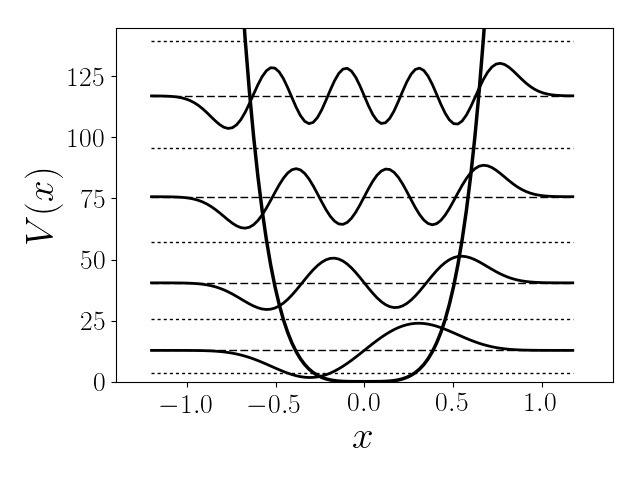}} 
\caption{The four odd eigenfunctions (narrow solid lines) found analytically for $N = 3$, with $V_0=128$, together with the corresponding eigenvalues (dashed lines).  The unsolved even eigenvalues are shown in dotted lines.}
\label{figA3}
\end{figure}

\begin{equation}
\begin{matrix}
\psi_{3,1}(x) = &\mathrm{senh} \left( x \right) e^{-8 \cosh^{2} \left( x \right)} \left( \cosh^{2} \left( x \right) - 0.3054 \right) \left( \cosh^{2} \left( x \right) - 0.1000 \right) 
\left( \cosh^{2} \left( x \right) - 0.0107 \right) \\ 

\psi_{3,3}(x) = &\mathrm{senh} \left( x \right) e^{-8 \cosh^{2} \left( x \right)} \left( \cosh^{2} \left( x \right) - 1.1278 \right) \left( \cosh^{2} \left( x \right) - 0.1380 \right)
\left( \cosh^{2} \left( x \right) - 0.0141 \right) \\

\psi_{3,5}(x) = &\mathrm{senh} \left( x \right) e^{-8 \cosh^{2} \left( x \right)} \left( \cosh^{2} \left( x \right) - 1.2916 \right) \left( \cosh^{2} \left( x \right) - 1.0666 \right)
\left( \cosh^{2} \left( x \right) - 0.0238 \right) \\

\psi_{3,7}(x) = &\mathrm{senh} \left( x \right) e^{-8 \cosh^{2} \left( x \right)} \left( \cosh^{2} \left( x \right) - 1.4497 \right) \left( \cosh^{2} \left( x \right) - 1.1795 \right)
\left( \cosh^{2} \left( x \right) - 1.0428 \right) 
\end{matrix}
\end{equation}
shown in Fig.(\ref{figA3}).

A summary of all even and odd energy eigenvalues found analytically, up to $N=4$, is given in Table \ref{tablaener}. All eigenvalues and eigenfunctions shapes were checked using Numerov's method.\cite{Paolo}


\begin{table}[H]
\begin{center}
\begin{tabular}{|l||l|l|c||l|l|c|}
\hline
$N$ & $V_0$ & $n$ & $E_{N,n}$ & $V_0$ & $n$ & $E_{N,n}$  \\
\hline
\hline

\multirow{1}{0.8cm}{$0$} & \multirow{1}{0.8cm}{$2$} & 
$0$ & $E^{S}_{0,0}=1.0000$ & \multirow{1}{0.8cm}{$8$} & $1$ & $E^{A}_{0,1}=5.5000$ \\  
\hline
\hline

\multirow{2}{0.8cm}{$1$} & \multirow{2}{0.8cm}{$18$} & 
$0$ & $E^{S}_{1,0}=1.9172$ & \multirow{2}{0.8cm}{$32$} & $1$ & $E^{A}_{1,1}=8.3348$ \\ 
\cline{3-4} \cline{6-7}
 & & 2 & $E^{S}_{1,2}=14.0828$ & & $3$ & $E^{A}_{1,3}=26.6652$ \\
\hline
\hline

\multirow{3}{0.8cm}{$2$} & \multirow{3}{0.8cm}{$50$} & 
$0$ & $E^{S}_{2,0}=2.6301$ & \multirow{3}{0.8cm}{$72$} & $1$ & $E^{A}_{2,1}=10.7038$ \\ 
\cline{3-4} \cline{6-7}
 & & $2$ & $E^{S}_{2,2}=19.0121$ & & $3$ & $E^{A}_{2,3}=33.9628$ \\ 
\cline{3-4} \cline{6-7}
 & & $4$ & $E^{S}_{2,4}=43.2490$ & & $5$ & $E^{A}_{2,5}=63.8335$ \\
\hline
\hline

\multirow{4}{0.8cm}{$3$} & \multirow{4}{0.8cm}{$98$} & 
$0$ & $E^{S}_{3,0}=3.2503$ & \multirow{4}{0.8cm}{$128$} & $1$ & $E^{A}_{3,1}=12.8152$ \\
\cline{3-4} \cline{6-7}
 & & $2$ & $E^{S}_{3,2}=23.4960$ & & $3$ & $E^{A}_{3,3}=40.4568$ \\
\cline{3-4} \cline{6-7}
 & & $4$ & $E^{S}_{3,4}=52.8353$ & & $5$ & $E^{A}_{3,5}=75.7246$ \\
\cline{3-4} \cline{6-7}
 & & $6$ & $E^{S}_{3,6}=88.4180$ & & $7$ & $E^{A}_{3,5}=117.0030$ \\
\hline
\hline

\multirow{5}{0.8cm}{$4$} & \multirow{5}{0.8cm}{$162$} & 
$0$ & $E^{S}_{4,0}=3.8129$ & \multirow{5}{0.8cm}{$200$} & $1$ & $E^{A}_{4,1}=14.7542$ \\
\cline{3-4} \cline{6-7}
 & & $2$ & $E^{S}_{4,2}=27.4612$ & & $3$ & $E^{A}_{4,3}=46.4155$ \\
\cline{3-4} \cline{6-7}
 & & $4$ & $E^{S}_{4,4}=61.5132$ & & $5$ & $E^{A}_{4,5}=86.6258$ \\
\cline{3-4} \cline{6-7}
 & & $6$ & $E^{S}_{4,6}=102.6240$ & & $7$ & $E^{A}_{4,7}=133.5300$ \\
\cline{3-4} \cline{6-7}
 & & $8$ & $E^{S}_{4,8}=149.5890$ & & $9$ & $E^{A}_{4,9}=186.1740$ \\
\hline
\end{tabular}
\caption{Energy eigenvalues for symmetric and antisymmetric solutions up to $N=4$, found using polynomial solutions to eqs.(\ref{ecuacion5.1}) and (\ref{ecuacion32}).} 
\label{tablaener}
\end{center}
\end{table}

It is possible to show that our even and odd solutions, for $k=0$, correspond to the solutions in Ref.\cite{no2int}, for the parameter pairs $(\sigma,\eta)=(1,0)$, and $(0,-1)$, respectively, with 
$\zeta=M$.  Also, we may insert an extra term $\beta^{1/2}=\cosh(x)$ in eqs.(\ref{ecuacion5.1},\ \ref{ecuacion32}), and end up with two CHEs which again possess polynomial solutions, which correspond to the other two combinations $(\sigma,\eta)=(-1,0)$, and $(0,1)$ in Ref.\cite{no2int}.  Moreover, these two new sets of solutions render the complementary odd/even part of the partial spectrum covered by the even/odd solutions found above.  For example, if one simply sets $f(\beta)=\beta^{1/2}$, in eq.(\ref{ecuacion5.1}), one gets $E_0=1.5$, which is the ground state eigenvalue for the potential with $V_0=8$, 
and if one sets $f(\beta)=\beta^{1/2}$ in eq.(\ref{ecuacion32}),
one gets $E_1=7$, which is the first excited state eigenvalue for the potential for with $V_0=18$, and so forth.
These are the first missing eigenvalues in Table \ref{tablaener} for the corresponding delimited spectrum of lowest $V_0$ values.


\section{The potential function 
$\boldsymbol{V(x) = V_0 \left( \mathrm{sinh}^{4}(x) - k \ \mathrm{sinh}^{2}(x)\right) }$}\label{seckgt0}

Now we can apply our previous analysis to the problem with the potential function 
$V(x) = V_0 \left( \mathrm{sinh}^{4}(x) - k \ \mathrm{sinh}^{2}(x)\right)$, which defines a symmetric double well when $k>0$.  

For the present case, to find even solutions we set again $\beta(x) = \cosh^{2} (x)$ and 
$\psi (\beta) = e^{- \frac{\alpha}{2} \beta} f (\beta)$, with $\alpha^2=2V_0$, rendering the following CHE
\begin{align}
\label{ecuacion5.5c5}
\nonumber \beta \left( \beta - 1 \right) \frac{d^{2} f}{d \beta^{2}} + &\left[ - \alpha \beta \left( \beta - 1 \right) + \left( \beta - \frac{1}{2} \right) \right] \frac{d f}{d \beta} \\
&+ \left[ \frac{\alpha^{2} \beta}{4} \left(1 + k \right) - \frac{\alpha \beta}{2} + \frac{\alpha}{4} + \frac{E}{2} - \frac{\alpha^{2}}{4} \left( 1 + k\right)\right] f = 0
\end{align}
with polynomial solutions. 

The solutions are found in like manner as in previous sections, and now $V_0=\frac{2(2N+1)^2}{1+k}$, $k$ varying freely. For example, in the even case with $N=0$, the only energy eigenvalue found is $E_{0,0}=1/(1+k)$, and no negative energy eigenvalues may be found analytically.

\begin{figure}[H]
\centering
{\includegraphics[width=0.7\textwidth]{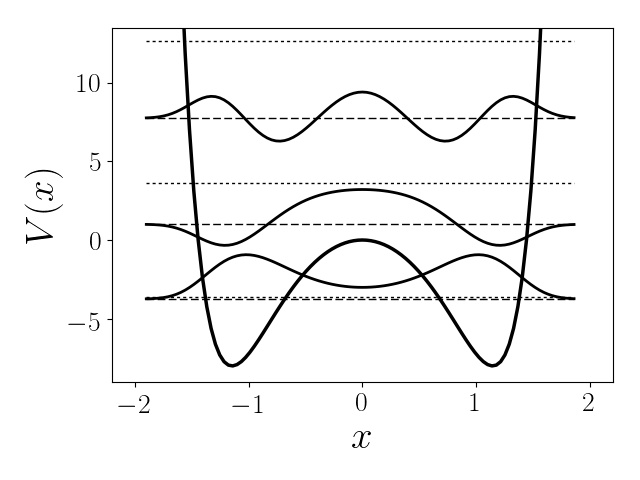}} 
\caption{The three even eigenfunctions (narrow solid lines) found analytically for $N = 2$, for the potential 
$V(x)=50\left({\rm sinh^4}(x)-4\,{\rm sinh^2}(x) \right)$ (bold solid line), together with the corresponding eigenvalues (dashed lines).  The unsolved odd eigenvalues are shown in dotted lines.}
\label{figN2k}
\end{figure}

Carrying on with our analysis, we can see that for $N=1$ the two energy eigenvalues found are
\begin{equation} \label{Ek-n1}
E = \frac{9 - \left( 1+ k \right) \pm \sqrt{ \left( 1+ k \right)^{2} + 36 }}{1+k}
\end{equation}
meaning that for $k>3/2$ we will have negative eigenvalues.  It is interesting to note that for $N>0$ it is always possible to find a value of $k$ which allows the appearance of a zero-energy groundstate, a feature that may have cosmological implications.\cite{socorro}

For the case with $N=2$, choosing $k=4$, the energy eigenvalues are 
$E_{2,0} = -3.74456$, $E_{2,2} = 1.00000$, and $E_{2,4} = 7.74456$.  The corresponding eigenfunctions are plotted in Fig.(\ref{figN2k}).

Now, if one wishes to find the antisymmetric eigenfunctions, we only need to use 
$f(\beta) = {\rm sinh}(x) \ g(\beta)$, to get the CHE
\begin{eqnarray} \label{ecuacion5.34}
\nonumber \beta (\beta - 1) \frac{d^{2} g}{d \beta^{2}} &+& \left[ -\alpha \beta^{2} + \left( \alpha + 2 \right) \beta - \frac{1}{2}   \right] \frac{d g}{d \beta} \\
&+& \left[\beta \left( \frac{\alpha^{2}}{4} \left( 1 + c \right) - \alpha \right) + \left( \frac{\alpha}{4}+ \frac{E}{2} - \frac{\alpha^{2}}{4}\left( 1 + c\right) + \frac{1}{4} \right) \right]g = 0
\end{eqnarray}
with polynomial solutions.

\begin{figure}[H]
\centering
{\includegraphics[width=0.7\textwidth]{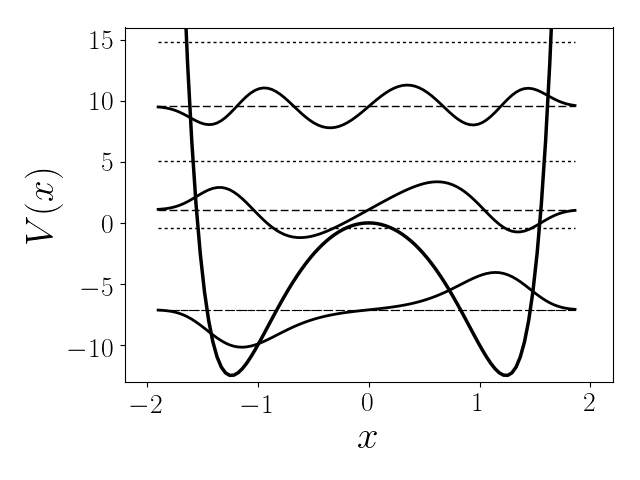}} 
\caption{The three odd eigenfunctions (narrow solid lines) found analytically for $N = 2$, for the potential 
$V(x)=2\left({\rm sinh^4}(x)-5\,{\rm sinh^2}(x) \right)$ (bold solid line), together with the corresponding eigenvalues (dashed lines).  The unsolved even eigenvalues are shown in dotted lines. } 
\label{figAN2k5}
\end{figure}

Here, if $N=0$ we get that $\alpha = 4/(1+k)$ and $E_1 = 6/(1+k) - 1/2$, such that if $k > 11$ we can assure finding negative energy eigenvalues.  As another example, for $N=2$ and $\alpha = 12/(1+k)$, if we set $k = 5$ 
the energy eigenvalues found analytically are 
$E_{2,1} = -7.11693$, $E_{2,3} = 1.08119$, and $E_{2,5} = 9.53574$.
The eigenfunctions are plotted in Fig.(\ref{figAN2k5}).

Note that in this case $(E_1-E_0)/E_0=0.0052$, and it is not possible to distinguish these eigenvalue's lines from each other in Fig.(\ref{figAN2k5}), implying quasi-degenerate eigenstates.  A similar effect is seen in Fig.(3).


\section{The case with $\boldsymbol{k=-1}$}\label{secnew}

As was seen in Section \ref{seckgt0}, the ground state energy diverges as $1/(1+k)$ as $k\to -1$, and this also happens to all higher order even eigenvalues (see eq.(\ref{Ek-n1})). This is a strange behaviour, since a plot of the potential function for any value of $k$ shows that it has a rather simple funtional form: a single or double well with infinite barriers.  

\begin{figure}[H]
\centering
{\includegraphics[width=0.7\textwidth]{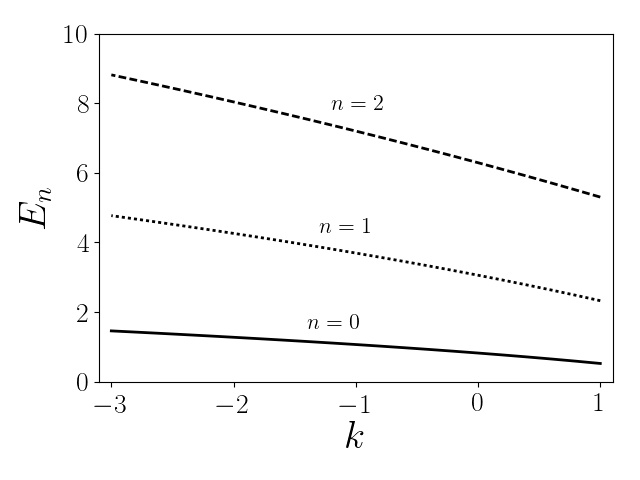}} 
\caption{The three lowest energy eigenvalues as a function of $k$ for $V_0=1$, showing that there are no singularities when $k\to -1$,but rather a smooth behaviour.}
\label{figkm1}
\end{figure}

We can see that this is only a charachteristic due to the analytical solution procedure: if we set
$V_0= 1$, a numerical analysis shows that there are no singularities when $k\to -1$, as can be seen in 
Fig.(\ref{figkm1}), where we let $k$ vary from -3 to 1.  
Note, however, that the energy eigenvalues seem to grow slower for smaller $k$, possibly reaching a constant value.

Note also that the potential strength $V_0$ is also divergent when $k\to -1$; moreover, it
diverges faster than the energy eigenvalues for the problem given in eq.(\ref{ecuaciondeschrodinger}), contradicting common sense, which would say that the larger $V_0$ the larger $E_n$.


\section{Concluding remarks}\label{final}

Here we have shown that the potential function 
$V(x) = V_{0} \left( \mathrm{sinh}^{4}(x)- k\ \mathrm{sinh}^{2}(x)\right)$ 
can be reduced to CHE's using the ansatz 
$\psi(x) \propto \sinh^q(x) \, e^{-\alpha\, {\rm cosh}^2(x)/2} \, f({\rm cosh}^2(x))$, where $q=0,1$ 
for even or odd solutions, and with $\alpha^2=2V_0$.
Since the CHE can be solved using finite polynomials, this potential corresponds to the class of QES potentials.
Also, it corresponds to the infinite well counterparts of the Maning potentials studied by 
Qiong-Tao.\cite{Qiong}

We have also shown that the singular eigenvalue bahaviour for $k\to -1$ is just an artifact of the analytical solution procedure, while this singular aspect is not present in the numerical solutions.



\begin{thebibliography}{999}

\bibitem{Turbiner} \textsc{Turbiner  A. V.}, Commun. Math. Phys. $\mathbf{118}$ (1988) 467.

\bibitem{Shifman} \textsc{Shifman M. A.}, Int. J. Mod. Phys. A $\mathbf{126}$ (1989) 2897.

\bibitem{Ushveridze1} \textsc{Ushveridze A. G.}, Sov. J. Part. Nucl. $\mathbf{20}$ (1989) 504.

\bibitem{Qiong} \textsc{Xie, Qiong-Tao}, J. Phys. A $\mathbf{45}$ (2012) 175302.

\bibitem{Razavy} \textsc{Razavy, M.}, Am. J. Phys. $\mathbf{48}$ (1980) 285-288.

\bibitem{Ushveridze2} \textsc{Ushveridze A. G.} \textit{Quasi-Exactly Solvable Models in Quantum Mechanics $\mathrm{Institute \ of \ Physics, \ Bristol, \ 1993}$}.

\bibitem{Chennn} \textsc{Chen, B.H}, {\it et al.}
J. Phys. A: Math. Theor. {\bf 46} (2013) 035301.

\bibitem{Oyewumi} \textsc{Oyewumi K. J.} and \textsc{Akoshile C. O.}, Eur. Phys. J. A $\mathbf{4}$ (2010) 578.

\bibitem{Wei} \textsc{Wei G. F.} and \textsc{Liu X. Y.}, Phys. Scr. $\mathbf{78}$ (2008) 065009.

\bibitem{Xie} \textsc{Xie W. F.}, Commun. Theor. Phys. $\mathbf{46}$ (2006) 1101.

\bibitem{Downing} \textsc{Downing C.A.}, J. Math. Phys. {\bf 54} (2013) 072101.

\bibitem{Wen} \textsc{Wen F.K.}, \textsc{Yang Z.Y.}, \textsc{Liu C.}. \textsc{Yang W-L} and \textsc{Zhang Y.Z.}, 
Commun. Theor. Phys. $\mathbf{61}$ (2014) 153-159.

\bibitem{Jaick} \textsc{El-Kaick, E.} and \textsc{Figuereido, B.D.B.}, J. Phys {\bf A, 48} (2013) 085203. 


\bibitem{no2int} \textsc{Finkel F., Gonzalez-Lopez A.}, and \textsc{Rodriguez M.A.}, J. Phys. A (Math. Gen.), {\bf 32} (1999) 6821.

\bibitem{no2int2} \textsc{Khare A.} and \textsc{Mandal B.P.}, J. Math. Phys. {\bf 39} (1998) 3476.
\bibitem{no2int3} \textsc{Konwent H., Machnikowsky P., Magnuszelwski P.}, and \textsc{Radosz A.}, Phys. Lett {\bf A, 31} (1998) 7541.

\bibitem{Yao} \textsc{Zhang Y.}, J. Phys. A $\mathbf{45}$ (2012) 065206.

\bibitem{edgar} \textsc{Condori-Pozo, E.}, Master's thesis at University of Guanajuato, Mexico, 2017.

\bibitem{socorro} \textsc{Socorro, J.} and \textsc{Nu\~{n}ez, O.M.},  Eur. Phys. J. Plus {\bf 132}: 168 (2017).

\bibitem{Paolo} \textsc{Giannozzi P.}, {\it``Numerical methods in quantum mechanics"}. Web address: \\
http://www.fisica.uniud.it/~giannozz/Corsi/MQ/mq.html$\,$.


\end{thebibliography}
\end{document}